\documentclass{Interspeech}

\usepackage{multirow}
\usepackage{tabularx}

\interspeechcameraready

\title{Segmentation-Variant Codebooks for Preservation of Paralinguistic and Prosodic Information}

\author{Nicholas}{Sanders}
\author{Yuanchao}{Li}
\author{Korin}{Richmond}
\author{Simon}{King}

\affiliation{The Centre for Speech Technology Research}{University of Edinburgh}{United Kingdom}
\email{\{nicholas.sanders, yuanchao.li, korin.richmond, simon.king\}@ed.ac.uk}

\usepackage{comment}

\begin{document}

\maketitle

\begin{abstract}
    

Quantization in SSL speech models (e.g., HuBERT) improves compression and performance in tasks like language modeling, resynthesis, and text-to-speech but often discards prosodic and paralinguistic information (e.g., emotion, prominence). While increasing codebook size mitigates some loss, it inefficiently raises bitrates. We propose Segmentation-Variant Codebooks (SVCs), which quantize speech at distinct linguistic units (frame, phone, word, utterance), factorizing it into multiple streams of segment-specific discrete features. Our results show that SVCs are significantly more effective at preserving prosodic and paralinguistic information across probing tasks. Additionally, we find that pooling before rather than after discretization better retains segment-level information. Resynthesis experiments further confirm improved style realization and slightly improved quality while preserving intelligibility.
\end{abstract}

\section{Introduction}
With the advent of pre-trained speech foundation models, significant advancements have been made in the field of speech processing. These models provide universal and comprehensive speech representations, leading to substantial improvements across various tasks, including Automatic Speech Recognition (ASR) \cite{baevski2020wav2vec, chang2024exploring}, Text-to-Speech (TTS) \cite{wang2023use, ulgen2024selectttssynthesizinganyonesvoice}, and Speech Emotion Recognition (SER) \cite{lilayer2022, chen2023wav2vecser}. However, as model sizes continue to grow, challenges arise in model deployment, data storage, and transmission efficiency. To address these challenges, quantization techniques have gained increasing attention, enabling the formation of Discrete Speech Units (DSUs) to improve computational efficiency.

DSUs, particularly those derived from Self-Supervised Learning (SSL) models, have proven highly effective in linguistically rich tasks such as speech recognition, translation, and understanding \cite{chang2024exploring}, as well as SER, which can heavily rely on lexical information \cite{saliba2024layer}. Additionally, DSUs have demonstrated better alignment with text tokens than continuous speech representations, making them advantageous for crossmodal ASR error correction \cite{li2024crossmodal} and speech large language models (LLMs) \cite{shon2024discreteslu}. This effectiveness can largely be attributed to two key factors: 1) SSL training objectives, such as masked segment prediction, which inherently focus on stable and predictable features like phonetic and lexical structures; and 2) KMeans clustering, which tends to group similar phonetic units together \cite{sanabria2023analyzing}.

However, while DSUs are capable of preserving linguistic information, they struggle to preserve paralinguistic and prosodic information \cite{nguyen2023expresso, ren2024emo}, presenting a trade-off between intelligibility and expressivity. Since phonemes are localized within short speech frames (e.g., 25 ms), SSL models inherently struggle to capture longer-span or sudden paralinguistic variations. Furthermore, the quantization process can lead to the loss of fine-grained or subtle paralinguistic details. For instance, while \textit{surprised} and \textit{angry} speech may share similar phonetic content (e.g., ``Are you kidding me?!''), they differ significantly in prosodic features such as pitch, intensity, and duration that DSU quantization often blurs.

Therefore, we pose a key question: \textit{Can DSUs be designed to better preserve paralinguistic and prosodic information?} To answer this, we make the following contributions in this work:
\begin{itemize}
    \item We propose \textbf{Segmentation-Variant Codebooks} (SVCs), a novel quantization approach that encodes speech at multiple linguistic levels (frame, phone, word, and utterance), effectively factorizing speech into distinct streams of segment-specific discrete units.
    \item We evaluate our approach against fixed frame-level DSUs and continuous speech representations across both probing tasks (SER and prominence classification) and resynthesis tasks, demonstrating improved expressivity while preserving speech intelligibility.
\end{itemize}

\section{Related Work}
Speech quantization facilitates efficient downstream processing by converting continuous speech representations into discrete, structured, and compact units. Although speech quantization methods do not have strictly defined categories, they can generally be classified into two major approaches: \textit{Vector Quantization (VQ)}, which is end-to-end trainable and continuously optimizes the codebook during training. Representative models include VQ-VAE \cite{van2017neural}, VQ-wav2vec \cite{baevski2020vq}, Encodec \cite{defossez2022high}, and SoundStream \cite{zeghidour2021soundstream}, which leverage VQ for low-bitrate speech compression and generation. \textit{Clustering-based quantization}, which applies statistical unsupervised clustering to learned speech representations, typically pre-training or post-training. KMeans clustering, used in HuBERT \cite{hsu2021hubert}, is a common approach for generating DSUs by grouping latent representations into phone-like clusters \cite{wells22_interspeech}.

Over the past few years, research on quantized speech representations has made significant progress, driven by the availability of large-scale unlabeled speech data as well as advanced neural networks and models. For instance: \cite{chang2024exploring} demonstrated that DSUs can achieve competitive performance in ASR and speech translation, even in low-resource settings. \cite{sicherman2023analysing} examined the interpretability of DSUs and their alignment with phonetic and linguistic structures. \cite{chang2024interspeech} introduced a unified evaluation platform to benchmark DSU performance across multiple speech tasks, including ASR, TTS, and singing voice synthesis.

However, while quantized speech effectively encodes phonetic and linguistic content, they face significant challenges in retaining paralinguistic and prosodic information. Since most DSUs are optimized for low-bitrate speech compression, they inherently prioritize phoneme-level representations and often fail to capture long-span paralinguistic variations.

To bridge this gap, we propose SVCs, with the aim to enhance discrete speech representations by capturing multi-level contextual dependencies and improving the preservation of paralinguistic and prosodic information with minimal compromise to linguistic integrity.

\section{Methodology}
\subsection{Segmentation-Variant Codebooks}
We encode speech into continuous representations using the frozen HuBERT-large model \cite{hsu2021hubert}. Our method operates as follows: first, speech inputs are encoded into frame-wise continuous representations via HuBERT. These representations are then pooled across segmentation boundaries (frames, phones, words, and utterances) derived from forced alignment applied to paired speech-text data. For phone-, word-, and utterance-level segments, mean pooling is applied to aggregate frame-wise representations within each segment, while frame-level representations remain unpooled. Each pooled or unpooled representation (corresponding to a segmentation level) is quantized using a dedicated Segmentation-Variant Codebook. To construct these codebooks, we train separate KMeans models (initialized with KMeans++) on the representations for each segmentation level: the frame-level codebook is trained on raw frame-wise HuBERT outputs, while phone-, word-, and utterance-level codebooks are trained on their respective pooled representations. During inference, quantization is performed by assigning each representation (pooled or unpooled) to the nearest cluster centroid in its corresponding codebook using Euclidean distance. This results in four parallel DSU output streams, each capturing linguistic structure at a distinct granularity (frame, phone, word, utterance).

\begin{figure*}[t]
  \centering
  \includegraphics[width=0.85\textwidth]{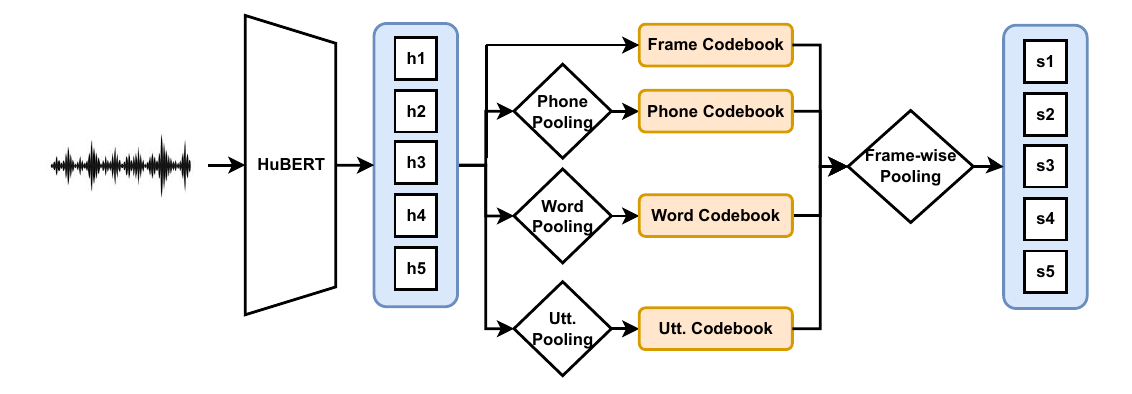}
  \caption{Segmentation-Variant Codebooks processing with pre-pooling overview. $h_n$ refers to continuous hidden representations at the $n$th frame, $S_n$ refers to the resulting frame-level stream obtained from mean pooling multiple streams of DSUs based on overlapping segmentation.}
  \label{fig:svc-overview}
\end{figure*}
\subsection{Processing Multiple Streams of DSUs}
For downstream tasks, there are potentially many ways that Segmentation-Variant DSU output streams could be processed. We choose to mean pool all DSUs across matching segmentations previously use to pool the continuous representations. For example, all frame DSUs that fall within the same utterance segment are averaged with the utterance DSU, all frame DSUs that fall within the same word segment are averaged with the same word DSU, and the same with the phones. Therefore, the resulting sequence is the same length as the frame-level stream of DSUs. However, we recognize that future work may want to explore other processing methods of the multiple streams of DSUs. The full method overview can be seen in Figure \ref{fig:svc-overview}.
\subsection{Bitrate Calculation}
We follow the approach described in \cite{chang2024interspeech} to calculate the bitrate $B$. The bitrate is defined as the total number of discrete units $N_m$ across $M$ streams, weighted by the base-2 logarithm of the vocabulary size $\lvert V_m \rvert$, divided by the duration of the sample $T$ scaled by the number of samples $S$. This can be expressed mathematically as
\begin{align}
    B = \sum_{m=1}^M \frac{N_m \cdot \log_2(|V_m|)}{T / S}
\end{align}
Because some $M$ streams are based on linguistic segmentations (e.g., word boundaries), the $N_m$ number of discrete units may vary from utterance to utterance. Therefore, in such cases we report an average bitrate for specific corpora. 

\begin{table}[h]
  \caption{Bitrate across datasets for Segmentation-Variant Codebooks (k = 500 for all codebooks) in comparison to other non-variable frame-wise baselines. NPC: Naver Prosody Control; Exp: Expresso.}
  \label{tab:average_bitrate}
  \centering
  \begin{tabular}{ l  r  r  r } 
    \toprule
    \textbf{Input Type} & \textbf{IEMOCAP} & \textbf{NPC} & \textbf{Exp} \\ 
    \midrule
    SVCs & 509.00 & 518.15 & 544.94 \\  
    Frames k=500        & \multicolumn{3}{c}{448.30} \\  
    Frames k=2000        & \multicolumn{3}{c}{548.30} \\  
    \bottomrule
  \end{tabular}
\end{table}

As shown in Table \ref{tab:average_bitrate} SVC results in a varied bitrate that is estimated for each dataset we use. This is especially in contrast to the other frame-level baselines we compare our method to.

\section{Experimental Conditions}
\subsection{Datasets and Alignment Process}
\textit{Naver Prosody Control} \cite{latif-etal-2021-controlling} is a dataset designed for studying prosody control in TTS systems. It provides spoken utterances with prosodic variations, particularly focusing on contrastive focus generation, making it suitable for prominence detection and prosody modeling. \textit{Expresso} \cite{nguyen2023expresso} is a benchmark dataset for expressive speech resynthesis. It includes high-quality recordings with diverse speaking styles and prosodic variations, allowing for the evaluation of speech synthesis models in terms of expressivity, style preservation, and prosody retention. \textit{IEMOCAP} \cite{busso2008iemocap} is a widely used SER dataset featuring acted and spontaneous dyadic conversations. It includes speech recordings annotated with categorical emotions (e.g., happy, sad, angry, neutral).

Forced alignments are obtained using the Montreal Forced Aligner \cite{mcauliffe17_interspeech} for Naver Prosody Control and Expresso, while IEMOCAP alignments are generated using the Hidden Markov Model Toolkit \cite{young2002htk}. Discretization for all baselines and SVCs is performed using the KMeans++ algorithm \cite{arthur2006k}, with codebooks trained exclusively on the respective training splits for each evaluation task. For instance, all codebooks used for SER evaluation on IEMOCAP are trained solely on the IEMOCAP training split. All SVCs are trained with a vocabulary of k = 500. 

\subsection{Probing Tasks}
We evaluate prominence classification on Naver Prosody Control and SER on IEMOCAP, using an 80-10-10 split for Naver Prosody Control and a session-based split for IEMOCAP (sessions 1–3 for training, 4 for validation, 5 for testing).

All probing tasks use a linear probe followed by softmax activation for SER and sigmoid activation for binary prominence classification. Optimization is performed using cross-entropy for SER and binary cross-entropy for prominence classification. Prominence is classified at either the word or frame level, while SER is conducted at either the utterance or frame level based on the input data (e.g., models trained on frame level representations make frame level predictions).

\subsection{Resynthesis}
For resynthesis, we follow a setup similar to the vocoder-only track in the Interspeech 2024 Challenge on Speech Processing Using Discrete Units \cite{chang2024interspeech}, training a modified HiFi-GAN \cite{kong2020hifi} on the Expresso dataset with its predefined splits. Representations are preprocessed and pooled outside the training process. For components without transcriptions, we use Whisper-large to transcribe conversational and long-form utterances. Other than adapting HiFi-GAN to accept HuBERT representations and conditioning on speaker labels, we make no model modifications. All models are trained for 100,000 steps with a batch size of 16.

\section{Experiments}
\subsection{Comparison of pooling pre-discretization and post-discretization}
In this study, we begin by comparing linear probing performance on features pooled before and after discretization to test whether discretization effectively factors out less salient prosodic and paralinguistic information. Expanding on previous successes in SER probing techniques \cite{yang2021superb, wang2021fine}, we hypothesize that aggregating continuous frame-wise representations prior to discretization, referred to as pre-pooling and shown in Figure \ref{fig:svc-overview}, preserves the more salient segment-level prosodic and paralinguistic features. Conversely, we suggest that performing discretization before pooling may lead to the loss of these important features due to the reduced saliency of paralinguistic and prosodic information within individual frame-wise representations. To investigate this hypothesis, we evaluated both SER at the utterance level and prominence classification at the word level, incorporating Segmentation-Variant codebook evaluations at the frame level, as demonstrated in \cite{wang2023speech, de2023emphassess}. This comprehensive analysis aims to determine the most effective pooling strategy for Segmentation-Variant representations.

\subsection{Performance comparison of frame-wise paralinguistic and prosody tasks}
Next, we assess the efficacy of SVCs in preserving paralinguistic and prosodic features by probing discrete frame-wise representations for prominence and emotion categories. Our benchmark for performance comparison utilizes continuous features as the top line, while we establish two discrete baselines using KMeans models with k = 500 and k = 2000. We hypothesize that our method will be more effective in maintaining paralinguistic and prosodic information, thereby outperforming these baselines, although it should underperform relative to the top line model trained with continuous representations.

\subsection{Resynthesis task for reconstructing expressive speech}
Lastly, we evaluate the performance of SVCs in a speech resynthesis task, quantifying expressive style retention through a publicly released pre-trained style classifier from Expresso. We measure intelligibility by calculating the Word Error Rate (WER) from Whisper's transcriptions of synthetic speech and compare these to the ground truth transcriptions of read speech. Additionally, we employ UTMOS scores \cite{saeki2022utmos} for overall quality evaluation, consistent with the approach taken in \cite{chang2024interspeech}. Ground truth waveforms provide the upper bounds for our objective metric comparisons. We hypothesize that our method will surpass the baselines of k = 500 and k = 2000 in terms of style classification while potentially trailing the continuous representations, reflecting the anticipated frame-wise probing results. Furthermore, we anticipate that our method will achieve intelligibility comparable to or exceeding that of the k = 2000 baseline, given that the bitrate of SVCs on the Expresso dataset is not significantly lower.

\section{Results}
\subsection{Factorization of paralinguistic and prosodic information via pooling}
For both prominence and emotion, Table \ref{tab:results-prepostpool} shows that pre-pooling is more effective than post-pooling, confirming our hypothesis. This effect is particularly pronounced for prominence classification compared to SER, possibly due to differences in task-specific segment lengths, though further investigation is needed. We also observe that this finding extends to our proposed method, SVCs, where probing DSUs is much more effective when pre-pooling. We interpret these results to mean that discretization results in the loss of prosodic and paralinguistic information, making post-pooling a less effective strategy than pre-pooling. This further underscores the advantage of utilizing multiple codebooks over a single one, as in post-pooling, for Segmentation-Variant conditioning. 

\subsection{Efficacy of preserving paralinguistic and prosodic information}
In Table \ref{tab:emotion_metrics} it can be observed that the use of SVCs outperforms both frame-level baslines of k=500 and k=2000 for every category in classification tasks. This confirms our hypothesis that SVCs are a much more effecient method of increasing bitrate, as the k=2000 baseline outperforms the k=500 baseline, but does not outperform SVCs despite being at a higher bitrate, as shown in Table \ref{tab:average_bitrate}. It is also worth noting that for many sub-categories, SVCs appears to approach the top-line performance of using continuous frame-level inputs, especially performing on par within reasonable probing-attributed errors for the Sad category. 

In Table \ref{tab:results-resynth} we can see that the style classifier performs best on the ground truth and that the model trained on continuous features approaches that level, however is still nearly 14\% off. The model trained with DSUs encoded by SVCs performs the best of the all models trained with DSUs. This further corroborates the results found in the frame-level probing experiments and further confirms our hypothesis that SVCs are more efficient at preserving paralinguistic and prosodic information. However, there is still a significant gap in performance between the model trained with DSUs encoded by SVCs and the model trained with continuous representations, further demonstrating that expressive resynthesis with DSUs is a challenging task. Audio samples are available\footnote{\href{https://yc-li20.github.io/Interspeech2025-SVC-audiosample/}{https://yc-li20.github.io/Interspeech2025-SVC-audiosample/}}.

\subsection{Acoustic quality and linguistic content preservation}
Although we did not anticipate a significant increase in linguistic content preservation or quality, the HiFi-GAN trained with SVCs encoded DSUs achieves lower WER and higher UTMOS than the other DSU-based models that were trained on k=500 and k=2000 frame-level codebooks. Overall, we observe that WER of Whisper transcribed ground truth speech is not as low as it may be in less-expressive data, such as LibriSpeech. We observe a similar effect with UTMOS, while also observing that the HiFi-GAN model trained with continuous representations achieves a similar WER as the ground truth but not a similar UTMOS. Finally, the results of all DSU-based resynthesis models fall short of the top-line model trained with continuous representations.

\begin{table}[t]
  \caption{Probing Accuracy Metrics of Comparing Pooling HuBERT Representations Post or Pre KMeans. Performance on IEMOCAP is reported as Accuracy of SER and performance on Naver Prosody is reported as Binary Fscore of Prominence Classification. The higher the score, the better.}
  \label{tab:results-prepostpool}
  \centering
  \begin{tabular}{ l  c  c  c  c }
    \toprule
    \textbf{Segmentation} & \multicolumn{2}{c}{\textbf{Emotion}} & \multicolumn{2}{c}{\textbf{Prominence}} \\
    \midrule
    & \textbf{Pre} & \textbf{Post} & \textbf{Pre} & \textbf{Post} \\

    \midrule
    Utterance & 0.5074  & 0.2834  & NA & NA \\
    Word      & NA     & NA     & 0.3423 & 0.1210 \\
    SVCs & 0.5699  & 0.4898  & 0.3050 & 0.1108 \\
    \bottomrule
  \end{tabular}
\end{table}

\begin{table}[t]
  \caption{Frame-level SER sub-class Micro F1 scores and prominence Binary F1 scores. The higher the score, the better.}
  \label{tab:emotion_metrics}
  \centering
  \begin{tabular}{l c c c c c}
    \toprule
    \textbf{Segmentation} & \textbf{Hap.} & \textbf{Sad} & \textbf{Ang.} & \textbf{Neut.} & \textbf{Prom.} \\
    \midrule
    k=500 & 0.070 & 0.304 & 0.272 & 0.429 & 0.115\\
    k=2000 & 0.045& 0.417 & 0.298 & 0.401 & 0.249 \\
    SVCs & 0.169 & 0.640 & 0.614 & 0.591 & 0.305 \\
    Continuous & 0.188 & 0.638 & 0.660 & 0.632 & 0.704\\

    \bottomrule
  \end{tabular}
\end{table}

\begin{table}[tp]
  \caption{Resynthesis Objective Results on Expresso. SVC refers to the model trained on Segmentation-Variant Codebook inputs. SCA refers to Style Classification Accuracy, WER is the Global Word Error Rate, and UTMOS is the predicted Mean Opinion Score. $\uparrow$: high the better. $\downarrow$: lower the better.}
  \label{tab:results-resynth}
  \centering
  \begin{tabular}{ l c c c }
    \toprule
    \textbf{Model} & \textbf{SCA$\uparrow$} & \textbf{WER$\downarrow$} & \textbf{UTMOS$\uparrow$} \\
    \midrule
    Ground Truth & 88.42\%  & 12.88\% & 3.4575 \\
    Continuous Features & 74.72\%  & 13.43\%  & 3.0759 \\
    Discrete Features k=500 & 24.53\%  & 20.72\%  & 2.2315 \\
    Discrete Features k=2000 & 31.63\%  & 18.57\%  & 2.3710 \\
    SVCs k=500 & 41.22\%  & 17.46\%  & 2.5135 \\
    \bottomrule
  \end{tabular}
\end{table}

\section{Discussion}
Our results demonstrate that SVCs efficiently preserve paralinguistic and prosodic speech qualities, as shown by both probing and resynthesis tasks. Performance may further improve by adjusting the codebook vocabulary size for better task adaptation.

While naively re-pooling DSUs encoded by SVCs may not be the most optimal strategy, our results show it still outperforms other frame-level baselines at similar or higher bitrates. This suggests that exploring more effective ways to process Segmentation-Variant DSUs could be a valuable direction for future DSU-based applications.

Additionally, the small but measurable gains in intelligibility and quality with SVCs warrant further investigation. Future studies should include human listening tests and qualitative analyses to identify potential error patterns. Another promising avenue is leveraging SVCs that integrate representations from different network layers, as intermediate features have been shown to enhance emotion \cite{lilayer2022} and stress \cite{de2024layer} classification.

Finally, a significant avenue for future work involves exploring alternative approaches to segmentation. While our current method relies on forced alignments, the inherent flexibility of SVCs could be further explored by investigating automatic and unsupervised segmentation methods \cite{sharma96_icslp}. Additionally, the promising evidence of factorization via pre-KMeans pooling of frame-wise representations suggests that further investigation is needed into how each Segmentation-Variant codebook contributes to different speech qualities, potentially informing targeted masking of certain codebooks based on downstream task requirements.

\section{Conclusion}
In conclusion, we propose Segmentation-Variant Codebooks, offering a promising approach for efficiently increasing representation power of DSUs by improving the preservation of paralinguistic and prosodic qualities. This study contributes to ongoing research on the use of DSUs, demonstrating their potential in speech representation learning and downstream tasks. Furthermore, our findings highlight promising future directions that could not only enhance the preservation of paralinguistic and prosodic features but also improve overall acoustic quality and linguistic content.

\begingroup
\renewcommand{\baselinestretch}{0.01} 
\small 
\section{Acknowledgments}
This work was supported in part by the UKRI CDT in NLP, funded by the UKRI (grant EP/S022481/1), the University of Edinburgh and Huawei.
\endgroup

\begingroup
\renewcommand{\baselinestretch}{0.01} 
\small 

\endgroup


\begin{thebibliography}{10}
\providecommand{\url}[1]{#1}
\csname url@samestyle\endcsname
\providecommand{\newblock}{\relax}
\providecommand{\bibinfo}[2]{#2}
\providecommand{\BIBentrySTDinterwordspacing}{\spaceskip=0pt\relax}
\providecommand{\BIBentryALTinterwordstretchfactor}{4}
\providecommand{\BIBentryALTinterwordspacing}{\spaceskip=\fontdimen2\font plus
\BIBentryALTinterwordstretchfactor\fontdimen3\font minus \fontdimen4\font\relax}
\providecommand{\BIBforeignlanguage}[2]{{%
\expandafter\ifx\csname l@#1\endcsname\relax
\typeout{** WARNING: IEEEtran.bst: No hyphenation pattern has been}%
\typeout{** loaded for the language `#1'. Using the pattern for}%
\typeout{** the default language instead.}%
\else
\language=\csname l@#1\endcsname
\fi
#2}}
\providecommand{\BIBdecl}{\relax}
\BIBdecl

\bibitem{baevski2020wav2vec}
A.~Baevski, Y.~Zhou, A.~Mohamed, and M.~Auli, ``wav2vec 2.0: A framework for self-supervised learning of speech representations,'' \emph{Advances in neural information processing systems}, vol.~33, pp. 12\,449--12\,460, 2020.

\bibitem{chang2024exploring}
X.~Chang, B.~Yan, K.~Choi, J.-W. Jung, Y.~Lu, S.~Maiti, R.~Sharma, J.~Shi, J.~Tian, S.~Watanabe \emph{et~al.}, ``Exploring speech recognition, translation, and understanding with discrete speech units: A comparative study,'' in \emph{ICASSP 2024-2024 IEEE International Conference on Acoustics, Speech and Signal Processing (ICASSP)}.\hskip 1em plus 0.5em minus 0.4em\relax IEEE, 2024, pp. 11\,481--11\,485.

\bibitem{wang2023use}
S.~Wang, G.~E. Henter, J.~Gustafson, and E.~Szekely, ``On the use of self-supervised speech representations in spontaneous speech synthesis,'' in \emph{12th Speech Synthesis Workshop (SSW)}, 2023.

\bibitem{ulgen2024selectttssynthesizinganyonesvoice}
I.~R. Ulgen, S.~S. Chandra, J.~Lu, and B.~Sisman, ``Selecttts: Synthesizing anyone's voice via discrete unit-based frame selection,'' 2024.

\bibitem{lilayer2022}
Y.~Li, Y.~Mohamied, P.~Bell, and C.~Lai, ``Exploration of a self-supervised speech model: A study on emotional corpora,'' in \emph{2022 IEEE Spoken Language Technology Workshop (SLT)}, 2023, pp. 868--875.

\bibitem{chen2023wav2vecser}
L.-W. Chen and A.~Rudnicky, ``Exploring {Wav2vec} 2.0 fine tuning for improved speech emotion recognition,'' in \emph{ICASSP 2023 - 2023 IEEE International Conference on Acoustics, Speech and Signal Processing (ICASSP)}, 2023.

\bibitem{saliba2024layer}
A.~Saliba, Y.~Li, R.~Sanabria, and C.~Lai, ``Layer-wise analysis of self-supervised acoustic word embeddings: A study on speech emotion recognition,'' in \emph{ICASSP 2024-2024 IEEE International Conference on Acoustics, Speech and Signal Processing Workshops (ICASSPW)}.\hskip 1em plus 0.5em minus 0.4em\relax IEEE, 2024.

\bibitem{li2024crossmodal}
Y.~Li, P.~Chen, P.~Bell, and C.~Lai, ``Crossmodal {ASR} error correction with discrete speech units,'' in \emph{2024 IEEE Spoken Language Technology Workshop (SLT)}.\hskip 1em plus 0.5em minus 0.4em\relax IEEE, 2024.

\bibitem{shon2024discreteslu}
S.~Shon, K.~Kim, Y.-T. Hsu, P.~Sridhar, S.~Watanabe, and K.~Livescu, ``Discreteslu: A large language model with self-supervised discrete speech units for spoken language understanding,'' \emph{arXiv preprint arXiv:2406.09345}, 2024.

\bibitem{sanabria2023analyzing}
R.~Sanabria, H.~Tang, and S.~Goldwater, ``Analyzing acoustic word embeddings from pre-trained self-supervised speech models,'' in \emph{ICASSP 2023-2023 IEEE International Conference on Acoustics, Speech and Signal Processing (ICASSP)}.\hskip 1em plus 0.5em minus 0.4em\relax IEEE, 2023.

\bibitem{nguyen2023expresso}
T.~A. Nguyen, W.-N. Hsu, A.~d'Avirro, B.~Shi, I.~Gat, M.~Fazel-Zarani, T.~Remez, J.~Copet, G.~Synnaeve, M.~Hassid \emph{et~al.}, ``Expresso: A benchmark and analysis of discrete expressive speech resynthesis,'' in \emph{INTERSPEECH 2023}.\hskip 1em plus 0.5em minus 0.4em\relax ISCA, 2023, pp. 4823--4827.

\bibitem{ren2024emo}
W.~Ren, Y.-C. Lin, H.-C. Chou, H.~Wu, Y.-C. Wu, C.-C. Lee, H.-y. Lee, H.-M. Wang, and Y.~Tsao, ``Emo-codec: An in-depth look at emotion preservation capacity of legacy and neural codec models with subjective and objective evaluations,'' in \emph{2024 Asia Pacific Signal and Information Processing Association Annual Summit and Conference (APSIPA ASC)}.\hskip 1em plus 0.5em minus 0.4em\relax IEEE, 2024.

\bibitem{van2017neural}
A.~Van Den~Oord, O.~Vinyals \emph{et~al.}, ``Neural discrete representation learning,'' \emph{Advances in neural information processing systems}, vol.~30, 2017.

\bibitem{baevski2020vq}
A.~Baevski, S.~Schneider, and M.~Auli, ``vq-wav2vec: Self-supervised learning of discrete speech representations,'' in \emph{International Conference on Learning Representations}, 2020.

\bibitem{defossez2022high}
A.~D{\'e}fossez, J.~Copet, G.~Synnaeve, and Y.~Adi, ``High fidelity neural audio compression,'' \emph{arXiv preprint arXiv:2210.13438}, 2022.

\bibitem{zeghidour2021soundstream}
N.~Zeghidour, A.~Luebs, A.~Omran, J.~Skoglund, and M.~Tagliasacchi, ``Soundstream: An end-to-end neural audio codec,'' \emph{IEEE/ACM Transactions on Audio, Speech, and Language Processing}, vol.~30, pp. 495--507, 2021.

\bibitem{hsu2021hubert}
W.-N. Hsu, B.~Bolte, Y.-H.~H. Tsai, K.~Lakhotia, R.~Salakhutdinov, and A.~Mohamed, ``Hubert: Self-supervised speech representation learning by masked prediction of hidden units,'' \emph{IEEE/ACM transactions on audio, speech, and language processing}, vol.~29, pp. 3451--3460, 2021.

\bibitem{wells22_interspeech}
D.~Wells, H.~Tang, and K.~Richmond, ``Phonetic analysis of self-supervised representations of english speech,'' in \emph{Interspeech 2022}, 2022, pp. 3583--3587.

\bibitem{sicherman2023analysing}
A.~Sicherman and Y.~Adi, ``Analysing discrete self supervised speech representation for spoken language modeling,'' in \emph{ICASSP 2023-2023 IEEE International Conference on Acoustics, Speech and Signal Processing (ICASSP)}.\hskip 1em plus 0.5em minus 0.4em\relax IEEE, 2023.

\bibitem{chang2024interspeech}
X.~Chang, J.~Shi, J.~Tian, Y.~Wu, Y.~Tang, Y.~Wu, S.~Watanabe, Y.~Adi, X.~Chen, and Q.~Jin, ``The {Interspeech} 2024 challenge on speech processing using discrete units,'' \emph{arXiv preprint arXiv:2406.07725}, 2024.

\bibitem{latif-etal-2021-controlling}
S.~Latif, I.~Kim, I.~Calapodescu, and L.~Besacier, ``Controlling prosody in end-to-end {TTS}: A case study on contrastive focus generation,'' in \emph{Proceedings of the 25th Conference on Computational Natural Language Learning}, A.~Bisazza and O.~Abend, Eds.\hskip 1em plus 0.5em minus 0.4em\relax Online: Association for Computational Linguistics, Nov. 2021, pp. 544--551.

\bibitem{busso2008iemocap}
C.~Busso, M.~Bulut, C.-C. Lee, A.~Kazemzadeh, E.~Mower, S.~Kim, J.~N. Chang, S.~Lee, and S.~S. Narayanan, ``{IEMOCAP}: Interactive emotional dyadic motion capture database,'' \emph{Language resources and evaluation}, vol.~42, pp. 335--359, 2008.

\bibitem{mcauliffe17_interspeech}
M.~McAuliffe, M.~Socolof, S.~Mihuc, M.~Wagner, and M.~Sonderegger, ``Montreal forced aligner: Trainable text-speech alignment using {Kaldi},'' in \emph{Interspeech 2017}, 2017, pp. 498--502.

\bibitem{young2002htk}
S.~Young, G.~Evermann, D.~Kershaw, G.~Moore, J.~Odell, D.~Ollason, V.~Valtchev, and P.~Woodland, ``The {HTK} book,'' \emph{Cambridge University Engineering Department}, vol.~3, 2002.

\bibitem{arthur2006k}
D.~Arthur and S.~Vassilvitskii, ``k-means++: The advantages of careful seeding,'' Stanford, Tech. Rep., 2006.

\bibitem{kong2020hifi}
J.~Kong, J.~Kim, and J.~Bae, ``Hifi-gan: Generative adversarial networks for efficient and high fidelity speech synthesis,'' \emph{Advances in neural information processing systems}, vol.~33, pp. 17\,022--17\,033, 2020.

\bibitem{yang2021superb}
S.~W. Yang, P.~H. Chi, Y.~S. Chuang, C.~I.~J. Lai, K.~Lakhotia, Y.~Y. Lin, A.~T. Liu, J.~Shi, X.~Chang, G.~T. Lin \emph{et~al.}, ``Superb: Speech processing universal performance benchmark,'' in \emph{22nd Annual Conference of the International Speech Communication Association, INTERSPEECH 2021}.\hskip 1em plus 0.5em minus 0.4em\relax International Speech Communication Association, 2021, pp. 3161--3165.

\bibitem{wang2021fine}
Y.~Wang, A.~Boumadane, and A.~Heba, ``A fine-tuned {Wav2vec} 2.0/{HuBERT} benchmark for speech emotion recognition, speaker verification and spoken language understanding,'' \emph{arXiv preprint arXiv:2111.02735}, 2021.

\bibitem{wang2023speech}
Y.~Wang, M.~Ravanelli, and A.~Yacoubi, ``Speech emotion diarization: Which emotion appears when?'' in \emph{2023 IEEE Automatic Speech Recognition and Understanding Workshop (ASRU)}.\hskip 1em plus 0.5em minus 0.4em\relax IEEE, 2023, pp. 1--7.

\bibitem{de2023emphassess}
M.~de~Seyssel, A.~D'Avirro, A.~Williams, and E.~Dupoux, ``Emphassess: a prosodic benchmark on assessing emphasis transfer in speech-to-speech models,'' \emph{arXiv preprint arXiv:2312.14069}, 2023.

\bibitem{saeki2022utmos}
T.~Saeki, D.~Xin, W.~Nakata, T.~Koriyama, S.~Takamichi, and H.~Saruwatari, ``{UTMOS}: {UTokyo-SaruLab} system for voicemos challenge 2022,'' in \emph{Proceedings of the Annual Conference of the International Speech Communication Association, INTERSPEECH}, vol. 2022, 2022, pp. 4521--4525.

\bibitem{de2024layer}
A.~de~la Fuente and D.~Jurafsky, ``A layer-wise analysis of mandarin and english suprasegmentals in ssl speech models,'' in \emph{Proc. Interspeech 2024}, 2024, pp. 1290--1294.

\bibitem{sharma96_icslp}
M.~Sharma and R.~J. Mammone, ``blind speech segmentation: automatic segmentation of speech without linguistic knowledge,'' in \emph{4th International Conference on Spoken Language Processing (ICSLP 1996)}, 1996, pp. 1237--1240.

\end{thebibliography}
\end{document}